\documentclass{appolb}
\usepackage{graphicx}

\usepackage{amsmath}
\usepackage[colorlinks=true]{hyperref}
\hypersetup{
    colorlinks=true,
    linkcolor=red,
    citecolor=blue,
    filecolor=magenta,      
    urlcolor=blue,
    }
  \usepackage[
    backend=biber,
    style=nature,
  ]{biblatex}

 \bibliography{references}

\begin{document}
\title{Shape coexistence and superdeformation in $^{28}$Si%
\thanks{Presented at the XXXVII Mazurian Lakes Conference on Physics, Piaski, Poland, September
3-9, 2023.}%
}
\author{D. Frycz, J. Menéndez$^{\dagger}$, A. Rios$^{\dagger}$, A. M. Romero$^{\dagger}$
\address{Departament de Física Quántica i Astrofísica, Universitat de Barcelona, Spain. \\
$^{\dagger}$Institut de Ciències del Cosmos, Universitat de Barcelona, Barcelona, Spain.}
}


\maketitle
\begin{abstract}
We study the shape coexistence of differently deformed states within $^{28}$Si using shell-model and beyond-mean-field techniques. Experimentally, $^{28}$Si exhibits shape coexistence between an oblate ground state and an excited prolate structure. The oblate rotational band is described well within the $sd$ shell using the USDB interaction. However, for the prolate band, a modification of this interaction is required, lowering the single-particle energy of the $1d_{3/2}$ orbit. Furthermore, we explore the possibility of a superdeformed configuration in $^{28}$Si. Our calculations, spanning both the $sd$ and $pf$ shells, rule out the existence of a superdeformed $0^+$ bandhead within an excitation energy range of 10-20 MeV.

\end{abstract}
  
\section{Introduction}
Interactions among protons and neutrons lead to collective phenomena within the atomic nucleus. One notable outcome is the potential for the nucleus to adopt a permanent non-spherical shape in its own frame of reference. Quadrupole deformations, representing the most common deviations from a spherical form, are frequently observed in such cases. Additionally, within a single nucleus, one can observe diverse shapes at different excitation energies, a phenomenon referred to as shape coexistence, which is prevalent throughout the nuclear chart \cite{Shape_coexistence_experimental_view}.

\begin{figure}[t]
\centerline{%
\includegraphics[width=0.85\linewidth]{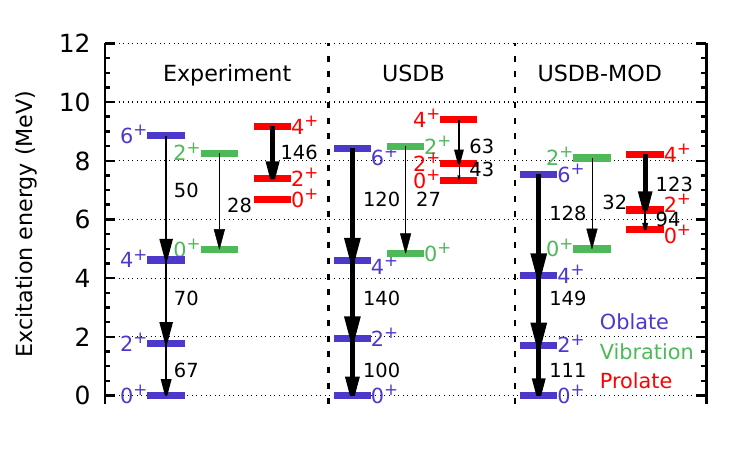}}
\caption{Band structure of the lowest-lying positive parity states of $^{28}$Si. Left: experimental data \cite{Si28exp}. Center: shell-model calculation using the USDB interaction. Right: same using USDB-MOD. The arrows indicate in-band $B(E2)$ transition strengths ($e$\(^2\) fm\(^4\)), larger values are associated to more deformed shapes.}
\label{Spectrum_Si28}
\end{figure}

According to experimental data, $^{28}$Si demonstrates shape coexistence \cite{Si28exp}. It exhibits an oblate rotational band built on its ground state, a vibration of the ground state with bandhead at $\sim5$ MeV, and a prolate rotational band with bandhead at $\sim6.5$ MeV, as shown in the left panel of Figure \ref{Spectrum_Si28}. Furthermore, there have been attempts at finding a superdeformed state \cite{Si28_SD}, proposed by a previous theoretical study \cite{AMD}. In this work, we study this shape coexistence with standard shell-model calculations and beyond mean-field techniques.

\section{Methods}
\begin{figure}[t]
\centerline{%
\includegraphics[width=0.45\linewidth]{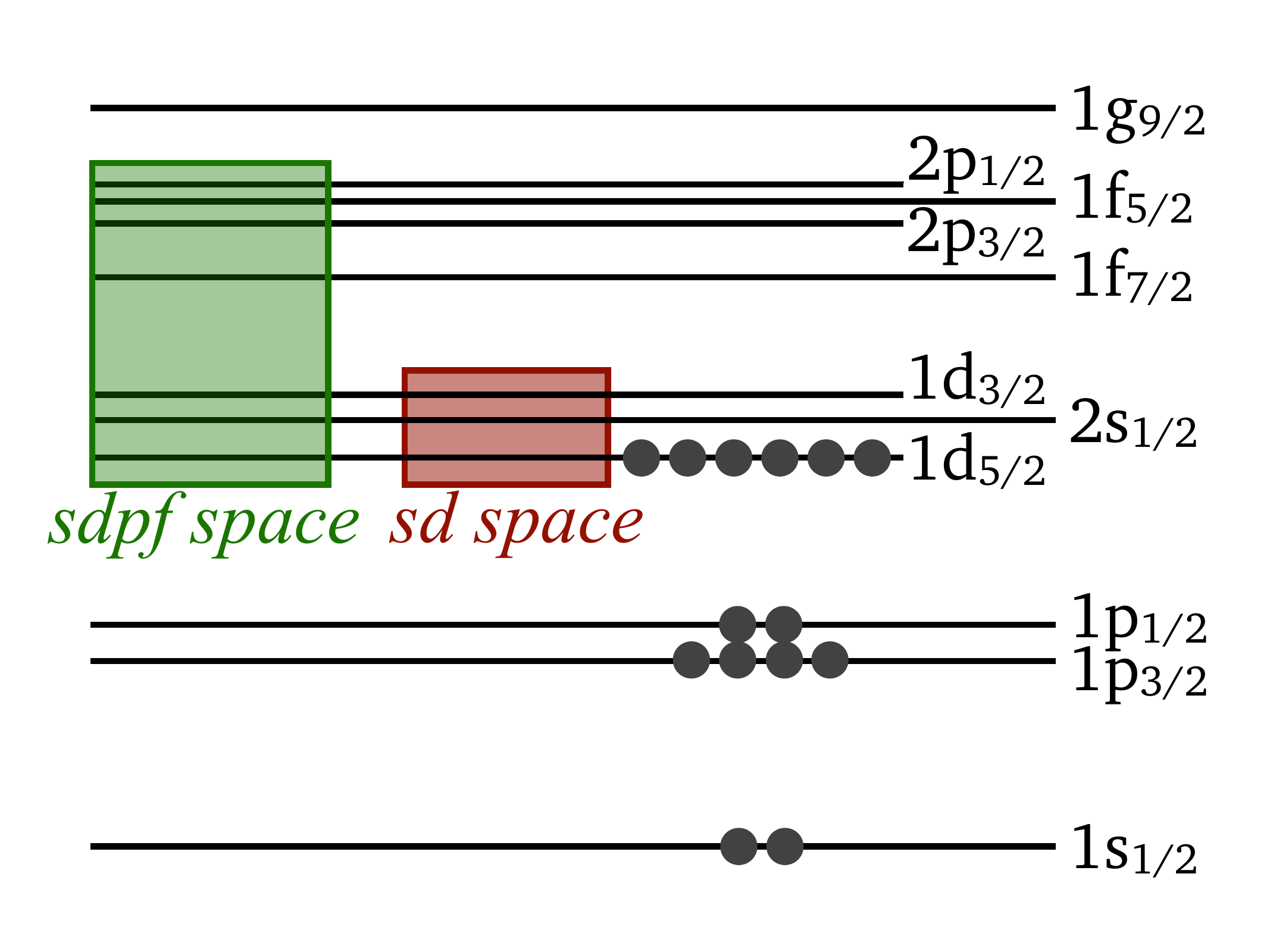}}
\caption{Standard spherical mean-field orbits. The circles represent the naive orbit filling for $^{28}$Si.}
\label{Shell_levels}
\end{figure}

{\bf Shell model:} We use the framework of the nuclear shell model \cite{Shellmodel}. A naive filling of the spherical mean-field levels, represented in Figure \ref{Shell_levels}, for $^{28}$Si with 14 protons and 14 neutrons results in the complete occupation of the $1d_{5/2}$ orbit. The associated Slater determinant is spherical, but the ground state of $^{28}$Si is experimentally known to be oblate.
Given that the Fermi level in the initial filling scheme resides within the $1d_{5/2}$ orbit, the $sd$ shell emerges as a natural choice for the valence space on an interacting nuclear shell model framework. Thus, we have an inert core of $^{16}$O, represented by a Slater determinant, along with 6 protons and 6 neutrons in the $sd$ shell. The effective interaction tailored for this valence space is USDB \cite{USDB}.

The task at hand entails the solution of the Schrödinger equation:
\begin{equation}
    \mathcal{H}\Psi=E\Psi,
\end{equation}
which reduces to the diagonalization of the Hamiltonian. The exact solution involves taking as a basis all of the Slater determinants $(\Phi)$ that can be built within the valence space. The configuration mixing of these basis states leads to the Hamiltonian eigenstates, which can be expressed as
\begin{equation}
    \Psi=\sum_{i}c_i \Phi_i.
\end{equation}
However, this methodology becomes unmanageable when the dimension of the valence space is large, as the required number of Slater determinants grows combinatorially with the available single-particle states. For instance, performing a calculation for $^{28} $Si encompassing both the $sd$ and $pf$ shells is unfeasible. 

{\bf Beyond mean-field approach:} An alternative approach comes from mean-field methods. Instead of a basis that consists of Slater determinants, we opt for a smaller set of more complex wavefunctions.
These states are found from variational methods, such as the Hartree-Fock-Bogoliubov (HFB) approach. To form the basis, the HFB wavefunctions are constrained to specific quadrupole momentum values $Q_{2\mu}=r^2Y_{2\mu}$, where $Y_{2\mu}$ are spherical harmonics. However, instead of constraining directly the quadrupole moments, we use the deformation parameters $(\beta,\gamma)$, defined as \cite{taurus}
\begin{equation}
    \beta=\frac{4\pi\sqrt{ {Q}_{20}^2+2\tilde{Q}_{22}^2}}{3r_0^{2}A^{5/3}}, \ 
    \gamma=\text{arctan}\left( \frac{\sqrt{2}\tilde{Q}_{22}}{ {Q}_{20}}\right), \tilde{Q}_{22}=\frac{(Q_{22}+Q_{2-2})}{2}
\end{equation}
with $r_0=1.2$ fm. The parameter $\beta$ characterizes the degree of deformation in the nucleus, with $\beta=0$ associated to a spherical shape. $\gamma$ characterizes the type of deformation, with $\gamma=0^{\circ}$ corresponding to prolate shapes and $\gamma=60^{\circ}$ corresponding to oblate shapes. 

With a set of quadrupole-constrained HFB wavefunctions $\phi(q)$, configuration mixing is performed using the generator-coordinate method (GCM)
\begin{eqnarray}
   \Psi = \sum_{q} f({q}) \hat{P}^{N} \hat{P}^{Z} \hat{P}^{J}   \phi(q) , 
\end{eqnarray}
where the HFB wavefunctions are projected onto good quantum numbers such as proton number $(Z)$, neutron number $(N)$ and total angular momentum $(J)$ employing operators $\hat{P}^{N}$, $\hat{P}^{Z}$ and $\hat{P}^{J}$ \cite{taurus}. This projection restores the broken symmetries at the mean-field level. Using this approach, implemented within the Taurus suite \cite{taurus}, we can investigate how the energy changes with different shapes of the nucleus and expand our analysis to larger valence spaces.

\section{Results}
\subsection{Oblate band}
\begin{figure}[b]
\centerline{%
\includegraphics[width=0.95\linewidth]{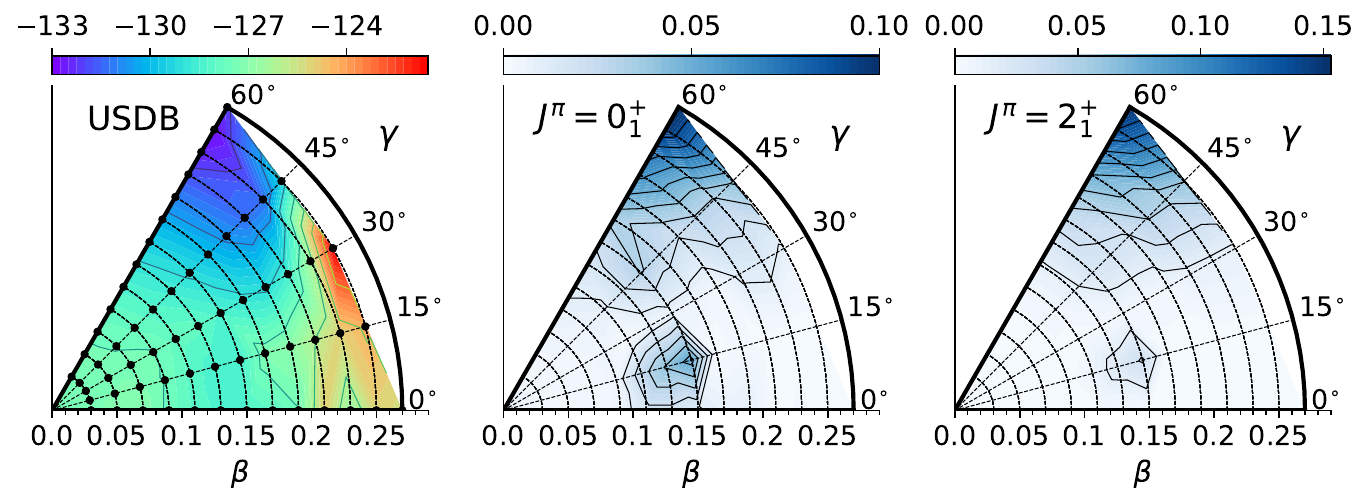}}
\caption{Projected total energy surface (MeV) obtained with the USDB interaction (left) and collective wavefunctions for oblate states with $J^{\pi}=0^+$ (center) and $J^{\pi}=2^+$ (right).}
\label{GCM_USDB}
\end{figure}

The ground state of $^{28}$Si can be estimated from the total energy surface of the projected HFB wavefunctions as a function of the deformation parameters, as represented in the left panel of Figure \ref{GCM_USDB}. The minimum energy for the ground state corresponds to an oblate shape with $\beta\simeq0.25$, in agreement with the deformation found experimentally. 

To obtain the nuclear states, we undertake configuration mixing through the GCM. The eigenstates resulting from this procedure are illustrated in the central panel of Figure \ref{Spectrum_Si28}. The oblate rotational band presents strong in-band $B(E2)$ transitions as well as an energy spacing proportional to $J(J+1)$. This indicates a well-established rotational band, although the states we find have larger deformation than the experimental ones. In addition, we can represent the weight of each HFB wavefunction to the mixed eigenstates with the collective wavefunctions \cite{taurus}. The two lowest states of the oblate rotational band are plotted in Figure \ref{GCM_USDB} (center and right panels). All states of the oblate band share a similar oblate deformation. Altogether, our calculations reproduce well the oblate band with the USDB interaction in the $sd$ shell. Additionally, we have checked that the GCM calculations are in excellent agreement with the exact diagonalization results obtained using the ANTOINE shell-model code \cite{Antoine,Antoine_2}.

\subsection{Prolate band}
\begin{figure}[b]
\centerline{%
\includegraphics[width=0.99\linewidth]{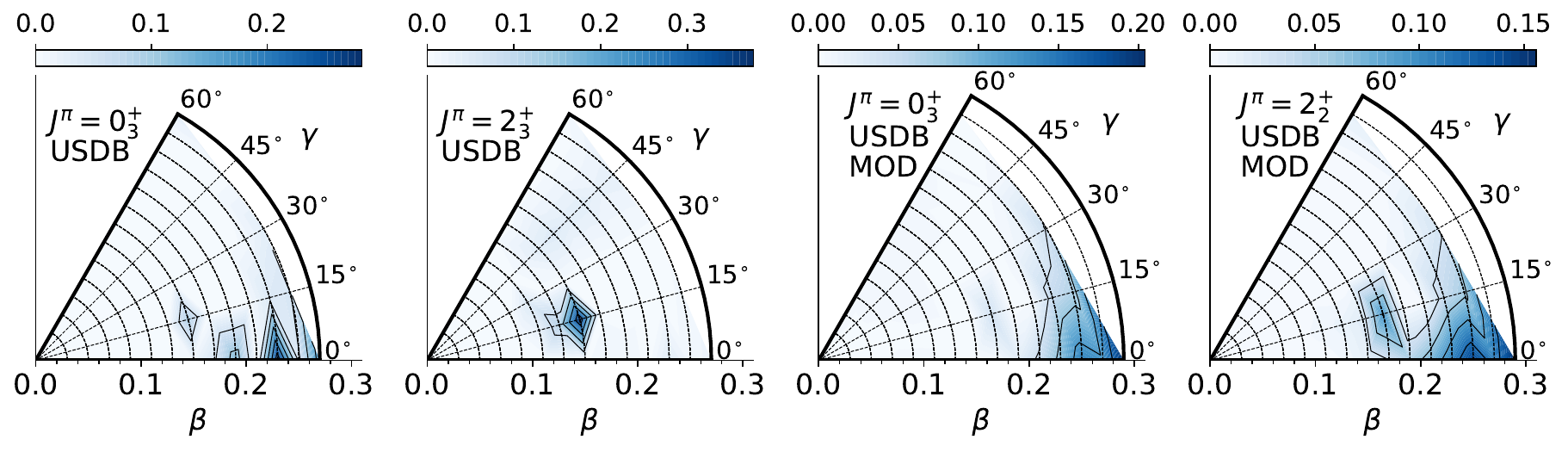}}
\caption{Collective wavefunctions for the prolate states with $J^{\pi}=0^+$ and $J^{\pi}=2^+$ for the USDB (two left panels) and USDB-MOD interactions (two right panels).}
\label{GCM_prolate}
\end{figure}

In contrast, when diagonalizing the USDB interaction within the $sd$ shell, our results do not reproduce well the prolate band. The states obtained exhibit weak in-band $B(E2)$ transitions and lack a consistent intrinsic deformation, as indicated by the collective wavefunctions shown in the two left panels of  Figure \ref{GCM_prolate}.

To accurately describe the prolate band, we slightly modify the USDB interaction. Employing a simple SU(3) scheme \cite{Elliot,Nilson}, the experimental deformation of the prolate band can be achieved by promoting particles from the closely situated $1d_{5/2}$ and $2s_{1/2}$ orbits to the $1d_{3/2}$ orbit, which for USDB has a $\sim5$ MeV gap. The reduction of this gap favours the promotion of particles to the $1d_{3/2}$ orbit. By reducing this energy gap by approximately $1.5$ MeV (USDB-MOD), a well-behaved rotational band consistent with experimental data emerges, while the oblate band remains largely unaffected. Notably, the prolate band states now exhibit a common deformation, as demonstrated in Figure \ref{GCM_prolate} (two right panels).

\subsection{Superdeformation}
In addition to the oblate and prolate bands, previous studies \cite{AMD} proposed a prolate superdeformed structure in $^{28}$Si with bandhead at $\sim14$ MeV. The structure would consist of promoting 4 particles from the $sd$ to the $pf$ shell with $\beta\approx0.6$. Nonetheless, recent experimental attempts \cite{Si28_SD} have not corroborated the existence of such a structure.

In order to reach such high deformations in our shell-model study, the $pf$ shell must be included in our calculations (green space in Fig. \ref{Shell_levels}). We employ the SDPF-NR interaction \cite{SDPF} to explore both major harmonic oscillator shells with the GCM method. We do not find any superdeformed structures $(\beta\ge0.5)$ within an energy range of $10-20$ MeV excitation energy, which disfavours the existence of a superdeformed band in $^{28}$Si.

\section{Summary}
In this study we describe the oblate rotational band of $^{28}$Si with shell model methods using the USDB interaction within the $sd$ shell. However, to reproduce the prolate band, we need to reduce the single-particle energy of the $d_{3/2}$ orbit. Additionally, we include the $pf$ shell to explore superdeformation. Our calculations do not find any superdeformed state within an excitation energy range of 10-20 MeV. \newline

This work is financially supported by 
grants PID2020-118758GB-I00 
funded by MCIN/AEI/10.13039/5011 00011033; 
by the ``Ram\'on y Cajal" grants RYC-2017-22781 and RYC2018-026072 funded by \\ MCIN/AEI/10.13039/50110 0011033 and FSE “El FSE invierte en tu futuro”;  
by the  “Unit of Excellence Mar\'ia de Maeztu 2020-2023” award to the Institute of Cosmos Sciences, Grant CEX2019-000918-M funded by MCIN/AEI/10.13039/501100011033; 
and by the Generalitat de Catalunya, grant 2021SGR01095.


\printbibliography[title={REFERENCES}]

\end{document}